\documentclass[%
pra%
,reprint%
,twocolumn%
,secnumarabic%
,floatfix%
,superscriptaddress
,showpacs%
,amssymb, amsmath, nobibnotes, aps]{revtex4-1}

\usepackage[latin1]{inputenc}
\usepackage{epsfig}
\usepackage{color}
\usepackage{comment}

\usepackage[colorlinks, linkcolor=red, citecolor=red, urlcolor=red, breaklinks]{hyperref}

\newcommand{\ind}[1]{_{\text{#1}}}

\newcommand{\mgbraket}[2]{\left.\left<#1\right|#2\right>}
\usepackage{braket}

\begin{document}

\title{Doublon dynamics of Bose-Fermi mixtures in optical lattices}

\author{Martin G\"arttner}
\affiliation{Kirchhoff-Institut f\"ur Physik, Universit\"at Heidelberg, Im Neuenheimer Feld 227, 69120 Heidelberg, Germany}
\affiliation{JILA, NIST and the University of Colorado, Boulder, Colorado 80309, USA}
\affiliation{Center for Theory of Quantum Matter, University of Colorado, Boulder, CO 80309, USA}
\author{Arghavan Safavi-Naini}
\affiliation{School of Mathematics and Physics, The University of Queensland, Brisbane, QLD 4072, Australia}
\affiliation{JILA, NIST and the University of Colorado, Boulder, Colorado 80309, USA}
\affiliation{Center for Theory of Quantum Matter, University of Colorado, Boulder, CO 80309, USA}
\author{Johannes Schachenmayer}
\affiliation{CNRS, IPCMS (UMR 7504), ISIS (UMR 7006), and Universit\'{e} de Strasbourg, 67000 Strasbourg, France}
\author{Ana Maria Rey}
\affiliation{JILA, NIST and the University of Colorado, Boulder, Colorado 80309, USA}
\affiliation{Center for Theory of Quantum Matter, University of Colorado, Boulder, CO 80309, USA}
\date{\today}

\begin{abstract}
We study the out-of-equilibrium dynamics of a dilute, lattice-confined Bose-Fermi mixture initialized in a highly excited state consisting of boson-fermion pairs (doublons) occupying single lattice sites. This system represents a paradigmatic case for studying relaxation dynamics in strongly correlated systems, and provides a versatile platform for studying thermalization and localization phenomena. We provide analytical expressions for the short-time decay of isolated doublons and small doublon clusters due to the competition between tunneling and interparticle interactions. We also discuss a mechanism for long-time decay that crucially depends on the quantum statistics of the particles constituting the doublon, namely, the conversion of pairs of neighboring doublons into an unpaired fermion and a site with a fermion and two bosons. Building on these insights, we develop a cluster expansion method to describe the dynamics in extended systems and compare it to numerically exact matrix product state simulations in one dimension. Finally, we discuss how our predictions can be observed in experiments with ultracold heteronuclear molecules.
\end{abstract}

\maketitle

\section{Introduction}
Understanding the relaxation dynamics of quantum systems out of equilibrium is vital to many outstanding problems in physics. In particular in the case of strongly interacting many-body systems and in the absence of a quasiparticle context, the study of such relaxation has profound consequences. Topics of interest range e.g.~from eigenstate thermalization in quantum statistical physics~\cite{Deutsch_Eigens_2018,Dalessio2016, Nandkishore_Many-B_2015} to fundamental questions in nuclear physics~\cite{Berges2018} or cosmology~\cite{Kofmann1997}. Due to the many-body nature of the problem, the theoretical analysis is very challenging and efficient numerical approaches are often restricted to one-dimensional (1D) systems~\cite{Orus_Aprac_2014,Schollwoeck2011,Verstraete_Matrix_2008}.
During the last decades it has become possible to realize strongly interacting closed quantum systems under highly controlled conditions using cold atoms trapped in optical lattices \cite{Bloch_Quantu_2018,Bloch_Quantu_2012,Lewenstein_Ultrac_2007}. 
In particular, experiments realizing tunable Fermi \cite{Joerdens2008,Esslinger_Fermi-_2010} and Bose-Hubbard models \cite{Greiner_Quantu_2002,Bakr2009} recently revealed a large range of interesting non-equilibrium phenomena such as many-body localization, strongly correlated multi-fermion scattering processes, or the direct observation of repulsively bound pairs (see \cite{Gross2017} for a recent review).

Hubbard models describing lattice-confined \emph{mixtures} of bosons and fermions are less well studied despite the fact that they also exhibit a rich phenomenology of relaxation dynamics. The early experimental efforts in trapping Bose-Fermi mixtures \cite{Gunter2006, Ospelkaus2006, Ospelkaus2006b} fueled theoretical investigations of \emph{equilibrium} properties of these mixtures, as well as those of the resulting dipolar bosonic or fermionic molecules into which the atom pairs are assembled \cite{Lang2008, Danzl2010}. For example, it was shown that Bose-Fermi mixtures can be used to stabilize a supersolid phase and charge density wave order, as well as quantum phases of composite fermions \cite{Titvinidze2008, Albus2003, Buchler2003, Lewenstein2004}. In addition, Bose-Fermi mixtures have been of interest in the context of cold molecule creation, where a number of groups has successfully created ultracold heteronuclear molecules \cite{Ospelkaus2006c, Demiranda2011, Reichsoellner2017, Seesselberg2018, Bohn2017}.

Here we study the dynamics in a Bose-Fermi mixture prepared in an out-of-equilibrium state in which bosons and fermions are paired into doublons on different sites. The dynamics of a {\em single} doublon after a quench have been studied extensively theoretically for Bose-Bose \cite{Winkler2006, HeckerDenschlag2006, Petrosyan2007, wang2008, Valiente2009, Javanainen2010, Deuchert2012}, Fermi-Fermi \cite{Valiente2010, Nguenang2009, Ohashi2008, Hofmann2012} and Bose-Fermi doublons \cite{Piil2008}. However, many aspects of the few and many-body physics of the doublons remain un-explored. We provide a  description of the decay dynamics of isolated and small clusters of doublons which will be essential for benchmarking the dynamics observed in quantum simulators. 

We find that in the case of Bose-Fermi doublons a decay channel involving the formation of triplons, i.e. sites occupied by a fermion and two bosons, facilitates the decay of doublons on neighboring sites -- a process which is not present for fermionic doublons. We study these processes and their importance on various time-scales using both analytical and numerical methods. We use our analytical results for small doublon clusters to efficiently calculate the short-time dynamics. We show that for the experimentally relevant case of strongly imbalanced tunneling rates the system can be understood in terms of the dynamics of highly mobile fermions in a lattice of defects formed by the heavier bosons. In the case where the inter-species interaction strength $U\ind{BF}$ is the dominant energy scale, the coupling between different energy bands separated by multiples of $U\ind{BF}$ can be treated perturbatively as illustrated in Fig.~\ref{fig:scheme}(b). In this limit, the decay of adjacent doublons into singlon-triplon pairs can also be understood in terms of a perturbative treatment. While the dynamics of Bose-Fermi mixtures in high dimensional extended geometries is in general numerically intractable, we show that in certain parameter regimes and on short time scales, cluster expansion techniques [see Fig.~\ref{fig:scheme}(a)] provide a good approximation to the full solution. We benchmark our approximate methods with exact t-DMRG simulations \cite{White2004, Daley2004, Schollwoeck2011} in 1d and using various analytical and numerical techniques. 

The paper is structured as follows: In section~\ref{sec:model} we introduce our model of Bose-Fermi mixtures in optical lattices. In section~\ref{sec:results} we present our results on single and few doublon dynamics and the cluster expansion model. Section~\ref{sec:conclusion} provides conclusions and an outlook on future directions, including a discussion of realistic experimental parameters for a possible implementation in a mixture of fermionic potassium and bosonic rubidium atoms.
Details of our analytical calculations are given in the appendix.

\section{Model}
\label{sec:model}

\begin{figure}[t]
  \centering
 \includegraphics[width=0.95\columnwidth]{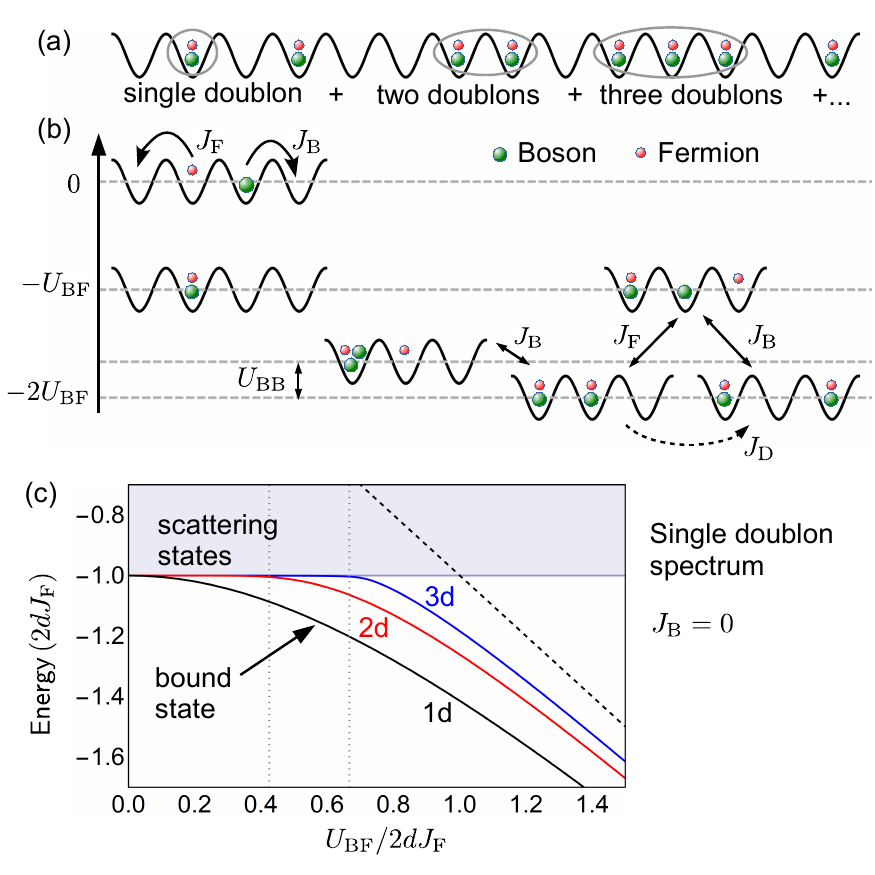}
 \caption{(a) Schematic illustration of the cluster expansion method: The initial state of doublons at random positions is partitioned into connected clusters. (b) Relevant states and parameters in the one and two doublon case. Both the bosons and the fermions are in the ground band of the optical lattice. Their tunneling rates are in general different for different masses and polarizabilities. (c) Eigenenergies in the single doublon case for $J\ind{B}=0$. The ground state (bound state) is shown as a black (red, blue) solid line for 1d (2d, 3d). Both axes have been scaled by half the with of the band of scattering states (blue shaded area) $2dJ\ind{F}$. In 2d and 3d there is no bound state below a certain critical interaction strength (gray dotted lines). The black dashed line indicates the ``bare'' bound state energy $U\ind{BF}$.}
 \label{fig:scheme}
\end{figure}

In the following we study the non-equilibrium dynamics of a Bose-Fermi mixture in a $d$-dimensional cubic lattice of linear size $L$. Atomic motion is restricted to the ground band of the lattice potential. The system is thus described by the single-band Hamiltonian
\begin{align}
\label{eq:HBF}
\nonumber H_{\rm BF}&= -J_{\rm B} \sum_{\langle \mathbf{i\, j }\rangle } a_{\mathbf i}^\dagger a_{\mathbf j} -J_{\rm F} \sum_{\langle \mathbf{i\, j }\rangle } c_{\mathbf i}^\dagger c_{\mathbf j} \\
&- U_{\rm BF} \sum_{\mathbf i} n_{\mathbf{i}, {\rm F}} n_{\mathbf{i}, {\rm B}} + U_{\rm BB}/2 \sum_{\mathbf i} n_{\mathbf{i}, {\rm B}} (n_{\mathbf{i}, {\rm B}}-1) 
\end{align}
where $a_{\mathbf{i}} (a_{\mathbf{i}}^\dagger)$ and $c_{\mathbf{i}} (c_{\mathbf{i}}^\dagger)$ are  bosonic and fermionic annhilation(creation) operators for a particle on the $d$-dimensional lattice site $\mathbf{i}$, respectively. The notation ${\langle \mathbf{i\, j }\rangle }$ indicates a summation over all neighboring sites and  $J_{\rm F}$ and $J_{\rm B}$ are the tunneling rates for the fermionic and bosonic species. The occupation number operator for a particle at site $\mathbf{i}$ is given by $n_{\mathbf{i},B}=a_{\mathbf i}^\dag a_{\mathbf i}$
($n_{\mathbf{i},F}=c_{\mathbf i}^\dag c_{\mathbf i}$).
$U_{\rm BB}$ denotes the on-site interaction strength between the bosonic particles, while $U_{\rm BF}$ is the inter-species on-site interaction. The individual terms of Hamiltonian~\eqref{eq:HBF} are shown schematically in Fig.~\ref{fig:scheme}(b).
Inspired by recent experiments with fermionic $^{40}$K and bosonic $^{87}$Rb atoms \cite{Covey2015}, we consider repulsive intra-species interactions and attractive inter-species interactions. We have defined the inter-species interaction term in the Hamiltonian with a negative sign such that both $U_{\rm BF}$ and $U_{\rm BB}$ take positive values.

The system is initially prepared in a product state where each site is either empty or occupied by a doublon, i.e.~a Bose-Fermi pair:
\begin{align}
\ket{\psi(t=0)}=\prod_{{\bf i}\in \mathcal{N}_{\rm occ}} a_{\mathbf i}^\dag c_{\mathbf i}^\dag \ket{\rm vac}.
\end{align}
Here $\mathcal{N}_{\rm occ}$ denotes the initially occupied sites, and $\ket{\rm vac}$ is the state of the empty lattice.
Our main observable is the normalized fraction of remaining doublons as function of time
\begin{align}
P_2(t)=\frac{1}{N} \sum_{{\bf i}} \bra{\psi(t)} p_{2,{\bf i}} \ket{\psi(t)}
\label{eq:init}
\end{align}
where $p_{2,{\bf i}}=a_{\bf i}^\dag c_{\bf i}^\dag \ket{\rm vac }\bra{ \rm vac} c_{\bf i} a_{\bf i}$ are projection operators onto local single doublon states. The sum runs over all $L^d$ sites of the lattice, and $\ket{\psi(t)}=\exp(- {\rm i} H t)\ket{\psi(0)}$ is the time-evolved state ($\hbar \equiv 1$). 
This choice of initial state and observable is motivated by recent experiments which have shown that both the preparation of an initial state of doublons and the detection of the doublon fraction is possible using magneto-association and dissociation of the doublons into Feshbach molecules \cite{Covey2015}, see also Sec.~\ref{sec:conclusion}.

In general, dynamics under $H\ind{BF}$ leads to a decay of $P_2(t)$. In the following we identify the main processes contributing to this decay by breaking down the full dynamics into dynamics of small clusters with single, two and three doublons [Fig.~\ref{fig:scheme}(a)].

\section{Results}
\label{sec:results} 

\subsection{Single doublon} 

\begin{figure}[t]
  \centering
 \includegraphics[width=\columnwidth]{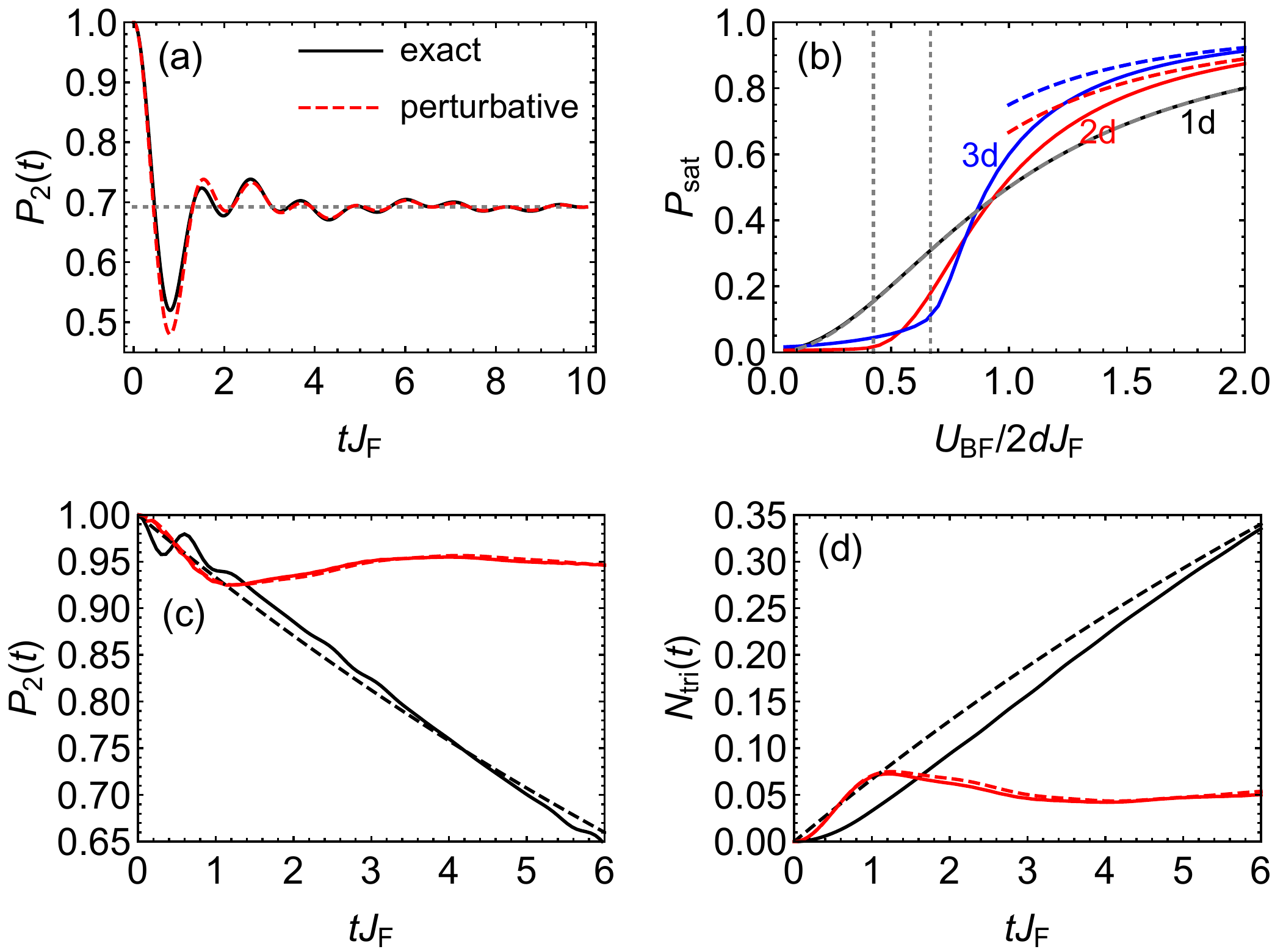}
 \caption{(a) Time evolution of the doublon fraction for a single-doublon initial state in 1d ($U\ind{BF}/J\ind{F}=3$, $J\ind{B}=0$). The dotted line indicates the saturation value $P\ind{sat}$. (b) Dependence of the doublon fraction on the inter-species interaction strength. Dashed lines: analytical prediction for $P\ind{sat}$ from perturbation theory. Linear system sizes for the $(1d,2d,3d)$ cases are $L=500,40,13$, respectively. In 1d (for $J\ind{B}$=0) the predicted saturation value is exact. The vertical dashed lines indicate the positions of $U_c$ in 2d and 3d, below which no bound state exists [compare Fig.~\ref{fig:scheme}(c)]. Note that this point is not defined sharply due to the finite system size. (c) Time evolution of the doublon fraction for initially two doublons on neighboring sites. The cases of two representative parameter sets are shown. Red (upper) and black (lower) solid lines are exact numerical results for these two parameter choices. The dashed lines are results of the analytical formulas in 1d, i.e. perturbative treatment (red, upper) and quasi continuum model (black, lower). (d) Time evolution of the triplon number. For the analytical results (dashed lines) our assumptions imply that $N\ind{tri}(t)=1-P_2(t)$. We used a 1d lattice of length $L=14$ with the two doublons initially on the two central sites. Parameters are: Case 1 (black, upper): $J\ind{B}/J\ind{F}=0.1$,$U\ind{BF}/J\ind{F}=-10$, $U\ind{BB}/J\ind{F}=1$; and case 2 (red, lower):  $J\ind{B}/J\ind{F}=0.2$,$U\ind{BF}/J\ind{F}=-30$, $U\ind{BB}/J\ind{F}=3$.}
 \label{fig:one_two_doublon}
\end{figure}

The decay dynamics of a heteronuclear doublon are governed by the relative magnitudes of $J_{\rm B}$, $J_{\rm F}$ and $U_{\rm BF}$, as well as the dimensionality of the system. Qualitatively, for large $U_{\rm BF}$ ($U\ind{BF}\gg J_{\rm B},J_{\rm F}$) doublons are bound and their decay is suppressed. For smaller $U_{\rm BF}$ the dimensionality of the system plays an important role. In contrast to 1d systems which support a bound-state for all $U_{\rm BF}\neq0$, 2d and 3d systems support a bound state only if $U_{\rm BF}>U_c$, with a critical interaction strength $U_c$, as shown in Fig~\ref{fig:scheme}(c). Hence for higher dimensional systems the doublon fraction decays to zero for $U_{\rm BF}<U_c$ and to a constant for larger attractive interactions. In 1d the problem of a single doublon can be solved analytically using the Greens function formalism described in ~\cite{HeckerDenschlag2006, Piil2008}. In the limit where $J_{\rm B}\ll J_{\rm F}$, i.e. when the boson tunneling is negligible, the problem further simplifies and one can use a perturbative approach to find a compact expression for the doublon fraction (see appendix \ref{sec:formalism} for details),  
\begin{equation}
\label{eq:DT1D} 
P_2(t) \approx P\ind{sat}\left[ 1 + \frac{4J_{\rm F} }{tU_{\rm BF}^2} J_1(2J_{\rm F} t)\cos\left(t\sqrt{4J_{\rm F} ^2+U_{\rm BF}^2}\right) \right]
\end{equation}
where $J_1(x)$ is the first Bessel function. It can be seen from the above expression that after initial oscillations at frequency $\sqrt{4J_{\rm F} ^2+U_{\rm BF}^2}$, the doublon fraction saturates to $P\ind{sat}=U_{\rm BF}^2/(U_{\rm BF}^2+4J_{\rm F}^2)$ after a time $t\ind{sat}\sim 1/(2J_{\rm F})$, see Fig.~\ref{fig:one_two_doublon}(a). This means that for $U\ind{BF} \gg J\ind{F}$, the doublon fraction saturates to a non-zero value while for $U\ind{BF}\lesssim 2J\ind{F}$ a significant doublon decay is observed as shown in Fig.~\ref{fig:one_two_doublon}(b). This is expected from a spectral analysis of the relevant state space shown in Fig.~\ref{fig:scheme}(c). Dividing the relevant states into bound states with $E_{\rm B} < -2dJ_{\rm F}$, connected to the initial state in the limit $J\ind{F}\rightarrow 0 $, and scattering states, we expect that these two manifolds become strongly mixed as soon as the bare bound state energy $U\ind{BF}$ enters the band of scattering states, which happens at $ U\ind{BF}=2J\ind{F}$ in 1d. The expression \eqref{eq:DT1D} can be generalized to the case of finite $J\ind{B}$ by separating the problem into center-of-mass and relative motion before applying the perturbative treatment to the relative motion part, see appendix~\ref{sec:formalism}. In higher dimensions it is not possible to obtain an analytical expression for $P_2(t)$. However, in the limit $J_{\rm F}\ll U_{\rm BF}$ one can obtain a good approximation for $P\ind{sat}$ by replacing $J\ind{F}$ by $\sqrt{d}J\ind{F}$ in the above expression. The statement that strong doublon decay occurs as soon as bound and scattering states overlap energetically ($U\ind{BF}\lesssim 2dJ\ind{F}$) still holds.

In the regime where $J_{\rm F}\gg J_{\rm B}$ the short-time behavior of systems of many doublons will be dominated by single doublon decay, as we will see when discussing the cluster expansion approach. This is consistent with the experimental observations made in~\cite{Covey2015}. The following analysis goes beyond this regime and takes into account additional decay channels that open through interactions between doublons.

\subsection{Two and three doublons} 

The single doublon decay described in the previous section occurs at a rate $J_{\rm F}$ and determines the short-time dynamics of the system. In the case of two doublons two additional processes have to be considered: i) If two doublons initially occupy neighboring sites, the tunneling of the boson results in the production of a triplon (two bosons and one fermion) and a singlon (lone fermion), cf.\ Fig.~\ref{fig:scheme}(b). The singlon can subsequently move through the lattice freely. This process occurs at a rate $\sim J\ind{B}$. ii) Another important indirect process is doublon tunneling to neighboring sites, which occurs at an effective tunneling rate $J_{\rm D}=2J_{\rm F} J_{\rm B}/U_{\rm BF}$. 

In the case that $U\ind{BF}$ is the largest scale in the problem, the doublon tunneling is slow compared to all other processes mentioned so far (see also the discussion below). Moreover, in the regime $J\ind{F}\ll U\ind{BF}$, the single doublon process of the fermion tunneling off the doublon site is strongly suppressed. Thus the process relevant to the decay of two neighboring doublons is the formation of the triplon-singlon complex and the subsequent tunneling of the singlon. 

Under these approximations, the dynamics is restricted to the initial two-doublon state and the singlon-triplon states. In order to find analytical expressions for the for the doublon fraction, we employ the perturbative approach used in the single-doublon case, supplemented by a quasi-continuum treatment of the scattering states (see appendix~\ref{sec:formalism} for details). Here, the two-doublon configuration takes the role of the bound state while the singlon plus triplon sector represents the scattering states. The energy splitting and coupling between the two sectors are given by $U\ind{BB}$ and $J\ind{B}$, respectively [see Fig.~\ref{fig:scheme}(b)]. The width of the band of scattering states is $4dJ\ind{F}$. Thus, in analogy with the single-doublon case, doublons are stable if the two sectors are energetically well separated ($U\ind{BB}\gg 2d J\ind{F}$) and decay if the bound and scattering states overlap ($U\ind{BB}< 2dJ\ind{F}$). In the former regime a perturbative treatment yields saturation value of $P\ind{sat}=U_{\rm BB}^2/(U_{\rm BB}^2+8J_{\rm B}^2)$ in 1d. In the latter regime, a quasi-continuum treatment of the scattering states shows that the doublon fraction decays exponentially with rate $\Gamma=8J\ind{B}^2/J\ind{F}\sqrt{1-[U\ind{BB}/(2J\ind{F})]^2}$ (see appendix \ref{sec:formalism}). In higher dimensions the behavior is qualitatively the same: If $U\ind{BB}\gg 2dJ\ind{F}$ two doublons on neighboring sites are essentially stable and $P_2(t)$ quickly saturates at a value near unity, while for $U\ind{BB}< 2dJ\ind{F}$, the doublon pair decays exponentially with a rate $\Gamma\propto J\ind{B}^2/J\ind{F}$. For $J\ind{B}\ll J\ind{F}$ the transition between both regimes is sharp.

Figures \ref{fig:one_two_doublon}(c) and (d) show the time evolution of the doublon and triplon fraction for two sets of parameters in 1d. This confirms that for $U\ind{BB}< 2dJ\ind{F}$ the doublon fraction decays quickly while for $U\ind{BB}> 2dJ\ind{F}$ it saturates. The triplon number increases in the same way the doublon fraction decreases, showing that the doublon decay is indeed caused by the formation of triplons. It should be noted that in our perturbative treatment the sum of the doublon fraction and the triplon fraction is always 1, as a direct result of the Hilbert space truncation. 

In the case of \emph{three doublons} on neighboring sites we observe that two among them can form a singlon plus triplon configuration, which leads to a decay of initially three doublons to one doublon in the regime $U\ind{BB}< 2dJ\ind{F}$ (see appendix \ref{sec:formalism} for details). 

\begin{figure}[tb]
  \centering
 \includegraphics[width=\columnwidth]{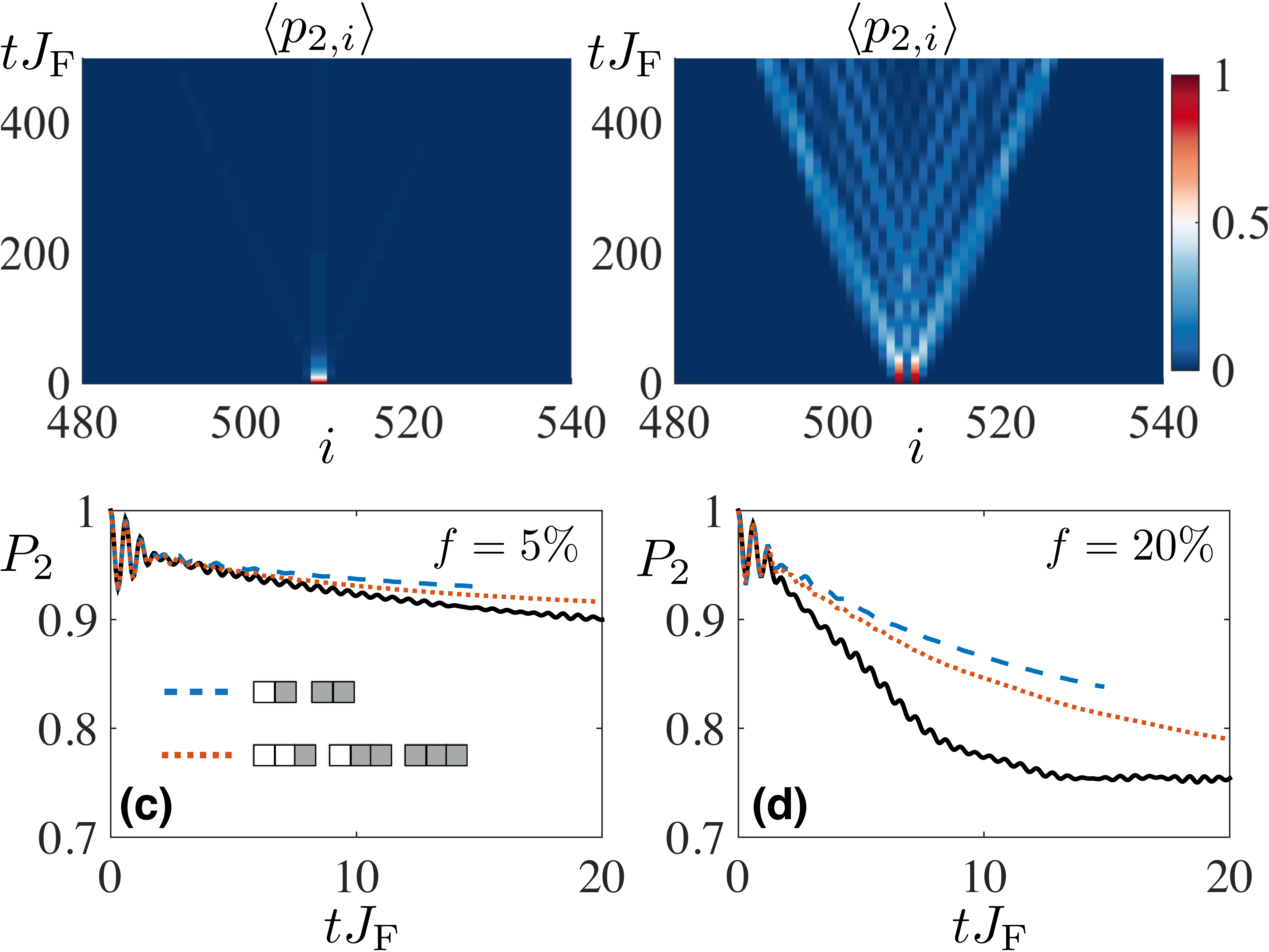}
 \caption{(a) and (b) Long-time evolution of the doublon density in a 1d lattice ($L=1010$) with two initial doublons computed using t-DMRG. Parameters are $J_{\rm B}= 0.1 J_{\rm F}$, $U_{\rm BB}=J_{\rm F}$, $U_{\rm BF}=10J_{\rm F}$. In (a) the initial doublons are located on neighboring sites, and there is a fast decay on a time-scale of $t \sim J_B^{-1} = 10 J_{\rm F}^{-1}$. In (b) the initial doublons are separated by an empty site. Here, the doublons are long-lived and diffuse on a time-scale of $t \sim J_{\rm D}^{-1}=100 J_{\rm F}^{-1}$. In this case the doublons do not decay by forming a triplon and singlon due to the quantum Zeno effect (see text). (c) and (d) Time-evolution of a system with $L=100$ sites and a random distribution of initial doublons with different filling fractions of $f=5\%$ and $f=20\%$ (panels c/d, respectively). For low fillings and short times, the exact t-DMRG solution (solid black line, average over $30$ initial configurations) is well reproduced by considering dynamics of small connected clusters. The dashed blue/dotted red line shows weighted averages of evolution of connected clusters with two/three doublons, respectively.}
 \label{fig:cluster_comp_overview}
\end{figure}

Having discussed the decay dynamics of a cluster of doublons initially sitting on neighboring sites we now discuss the case of two doublons separated by an empty site and $J_{\rm B}\ll J_{\rm F}\ll U_{\rm BF}$. This discussion will justify the use of only \emph{connected} clusters in our cluster expansion approach. In the limit $J_{\rm F}\ll U_{\rm BF}$ a single doublon is stable ($P_{\rm sat}\sim 1$), while two neighboring doublons may decay by forming a singlon-triplon configuration at rate $\Gamma$. Since doublon tunneling is strongly suppressed ($J\ind{D}\ll U_{\rm BF}$), the separated doublon pair is governed by the single doublon decay and thus stable. The doublons simply propagate on a time-scale given by $J_{\rm D}^{-1}$. This striking dependence on the initial doublon positions is confirmed in exact calculations shown in Figs.~\ref{fig:cluster_comp_overview}(a) and (b).

While one might expect to observe a decay process on a time scales $\sim J_{\rm D}^{-1}$ caused by decay after two doublons hopped on neighboring sites, this is not what we observe in the numerical computations of Fig.~\ref{fig:cluster_comp_overview}(b). The doublons remain stable in a much longer time scale than $J_{\rm D}^{-1}$. This can be understood by noting that $J_{\rm D}\ll \Gamma$, which places the system in a ``quantum Zeno regime''. In this regime, the coherent pair tunneling happens at a much slower rate compared to the decay process at rate $\Gamma$ which connects the two-doublon state to a continuum of scattering states consisting of a fixed triplon and a mobile singlon. As a result the decay process for two doublons initially separated by an empty site is suppressed. A perturbative expansion in $J\ind{D}/\Gamma$ shows that the effective doublon decay rate is given by $\Gamma_{\rm eff} \sim J_{\rm D}^2/\Gamma$. For the parameters considered in Fig.~\ref{fig:cluster_comp_overview}(b) this corresponds to a decay process occurring on a time-scale of $\sim 200 J_{\rm F}^{-1}$, consistent with our numerical observations. The discussed effects make it plausible why one may accurately describe the dynamics of a dilute system by considering solely the dynamics of \emph{connected} few-doublon clusters.

\subsection{Extended systems: Cluster expansion}

Next, we consider extended systems of randomly placed doublons at a given filling fraction.
An upper limit on the propagation speed of particles is given by the maximal group velocity of the most mobile particles, typically the fermions, $v_g=\partial E/\partial k\leq 2J\ind{F}$ (1d) limiting the speed at which correlations can spread in the system. Thus, in order to calculate the value of a local observable at site $i$ after time $t$ it is always sufficient to simulate a volume of radius $r=2J\ind{F}t$ around site $i$. If the system shows some kind of localization length $\xi<2J\ind{F}t$ a smaller simulation volume might be sufficient. For a translationally invariant system this is equivalent to simulating a box of linear size $L>2J\ind{F}t$ with periodic boundary conditions and evaluating the observable at all sites of the box. Thus short-time dynamics can always be simulated exactly.

As described above, we know that in the case of large $U\ind{BF}$ the decay rate of two initially separated doublons is small. Thus, in this case a viable ansatz is to consider only the dynamics of \emph{connected} clusters up to a certain size. This cutoff size can be small for low densities. For example, in 3d, at a filling fraction of $f=10\%$ the fraction of doublons that are part of a connected cluster of size larger than $3$ is about $1.5\%$. In this case the process dominating the doublon decay, is the triplon formation from nearest neighbor doublons. This cluster expansion is schematically illustrated in Fig~\ref{fig:scheme}(a). 

In Fig.~\ref{fig:cluster_comp_overview}(c) and (d), the validity of this phenomenology is illustrated. There we compare an exact t-DMRG calculation in a 1d box with $L=100$ sites and initial small filling fractions $f=5\%$ and $f=20\%$. To model the exact evolution we compare these results to weighted averages 
\begin{align}
\sum_{c=1}^{n_c} w_c P_2^{[c]}(t),
\end{align}
where  $P_2^{[c]}(t)$ denotes the evolution of a cluster with $c$ connected doublons and $w_c$ the  probability of the occurrence of the cluster in the initial state. Clearly this simple approach reproduces the exact calculation in the limit of short times and small filling fractions.

\section{Conclusions \& Outlook}
\label{sec:conclusion}

In conclusion we have studied the non-equilibrium dynamics of a lattice confined Bose-Fermi mixture initialized in a state consisting of doublons randomly placed on the lattice sites at a given dilute filling fraction. We found that the short-time relaxation dynamics can be understood in terms of the decay of connected doublon clusters. We have derived analytical expressions for the decay of clusters of one, two, and three doublons valid when the inter-species interaction is the dominating energy scale. We found that the relaxation of the doublon population is governed by multiple different decay time scales resulting from different processes such as single-doublon decay, doublon-pair decay via triplon formation, and decay after doublon tunneling at even longer times. We verified our cluster expansion model by numerically exact t-DMRG simulations in 1d.

Our choice of initial state and observable is motivated by the fact that both can be realized straight forwardly in experiment using the ability to associate and dissociate Feshbach molecules. In Ref.~\cite{Covey2015} a lattice gas of ground state KRb molecules was prepared where any remaining unpaired atoms can be removed from the lattice. After exciting the molecules to a weakly bound state (Feshbach molecules) they can subsequently be dissociated by an adiabatic ramp of the magnetic field, resulting in a configuration with all sites either empty of occupied by a doublon. The interspecies interaction strength can be adjusted by varying the endpoint of the magnetic field ramp. The remaining doublon fraction after a free evolution time can be determined by reversing the preparation process. Molecules only form on sites that are still occupied by a doublon. Unbound atoms can again be discarded and the molecules detected. Experiments that have taken advantage of the efficient conversions of pairs of the molecular constituents to molecules to create a low entropy lattice gas of polar molecules have recently been reported by a number of groups \cite{Moses2015, Takekoshi2012, Takekoshi2014, Park2012, Park2015, Grobner2015, Molony2014}, opening up a new direction for experimental investigation of \emph{non-equilibrium} dynamical properties of Bose-Fermi mixtures.

In order to show that the regimes studied above are indeed experimentally relevant, we provide typical parameters for the case of $^{40}$K and $^{87}$Rb atoms in an optical lattice of wavelength $\lambda=1064\,$nm as used in Ref.~\cite{Covey2015}. The different masses and polarizabilities lead to different tunneling rates for the two species. We express all parameters in units of $J\ind{F}=J\ind{K}$. $U\ind{BF}$ can be adjusted independently of the other parameters by varying the magnetic field near an inter-species Feshbach resonance. Increasing the lattice depth leads to slower tunneling and enhanced on-site interactions. Table \ref{tab:params} shows that all the different regimes are experimentally accessible, and also that $U\ind{BF}$ is indeed always the largest parameter and $J\ind{B}\ll J\ind{F}$ in this case. Thus a time resolved measurement of the doublon fraction should reveal the effects found here, and in addition provide a means to explore the long-time dynamics of large 3d ensembles, which are inaccessible to numerical simulations.

\begin{table}
\label{tab:params}
 \[\begin{array}{cccccc}
 \hline\hline
 V\ind{B}/E\ind{R} & a_s/a_0 & J\ind{B}/J\ind{F} & U\ind{BF}/J\ind{F} & U\ind{BB}/J\ind{F} & J\ind{D}/J\ind{F} \\
 \hline
 10 & -220 & 0.103 & 2.83 & 1.15 & 0.0728 \\
 10 & -910 & 0.103 & 11.7 & 1.15 & 0.0176 \\
 10 & -1900 & 0.103 & 24.5 & 1.15 & 0.00843 \\
 \hline
 15 & -220 & 0.059 & 7.05 & 2.81 & 0.0167 \\
 15 & -910 & 0.059 & 29.2 & 2.81 & 0.00405 \\
 15 & -1900 & 0.059 & 60.9 & 2.81 & 0.00194 \\
 \hline
 20 & -220 & 0.0372 & 15.2 & 5.94 & 0.00491 \\
 20 & -910 & 0.0372 & 62.7 & 5.94 & 0.00119 \\
 20 & -1900 & 0.0372 & 131. & 5.94 & 0.000569 \\ 
 \hline\hline
\end{array}\]
\caption{Typical experimental parameters for the case of fermionic $^{40}$K and bosonic $^{87}$Rb in a $\lambda=2\pi/k=1064\,$nm optical lattice. The experimentally adjustable parameters are the lattice depth $V\ind{B}$ (1st column, in units of lattice recoils of Rb, $E\ind{R}=\hbar^2k^2/2m\ind{Rb}$) and the interspecies scattering length (2nd column, in Bohr).}
\end{table}

The rich relaxation dynamics observed already for the restricted parameter regime considered here, and the prospect of experimental realization, call for a more systematic study of this system, addressing more parameter regimes and longer-time dynamics. Such a study would be quite challenging and require novel numerical and analytical methods for going beyond 1d and short times. Nevertheless, our results already showed a hierarchy of energy scales at which relaxation processes occur. Considering that such hierarchical relaxation is also found in spin glasses and soft matter physics, we take this as a hint for a rich phenomenolgy of relaxation dynamics in lattice Bose-Fermi mixtures yet to be uncovered. Adding an additional disorder potential, this system might be suited for studying localization phenomena in a new class of Hubbard models accessible to state-of-the-art quantum simulation experiments.

\appendix
\section{Details of analytical calculations of few-doublon dynamics}
\label{sec:formalism}

In this appendix we present details of the derivation of our analytical results for single, two and three-doublon clusters. In each scenario we calculate the doublon fraction, $P_2(t)$ from Eq.~\eqref{eq:init}.
For two and three doublons, the initial state is a connected cluster where the doublons are initialized on neighboring sites. 

\subsection{Single doublon physics}
\label{sec:from_single}

In this section we describe two approaches which yield analytic expressions for single doublon dynamics in 1d. These approaches will then be used to also obtain analytical results for more than one doublon within certain approximations. 

We first apply standard perturbation theory to the single doublon problem. The prescriptive nature of this method allows for straight forward extensions to larger number of doublons and higher dimensions. We start by considering the limit where one constituent of the doublon is stationary and can be approximated as a fixed defect. We then consider the case where both constituent atoms are mobile. Here we are able to build on the expressions derived in the prior approach and simply replace certain parameters.

\subsubsection{Perturbation Method: Stationary Defect}

Consider a system described by Hamiltonian~\eqref{eq:HBF} with $J\ind{B}=0$ on a ring of $L$ sites with periodic boundary conditions, where site $L$ is identified with site 0.
We use basis states $\lbrace \vert n \rangle \rbrace$, with $n=0, \dots, L-1$ corresponding to a  boson on site $i=0$ and the fermion on site $n$.
In this basis the system is described by the Hamiltonian
\begin{align}
 H&=H_0+H_1\\
 H_0&=-U_{\rm BF} \ket{0}\bra{0}+J_{\rm F} \sum_{n=1}^{L-2}(\ket{n}\bra{n+1}+\ket{n+1}\bra{n})\\
 H_1&=J_{\rm F} (\ket{0}\bra{1}+\ket{0}\bra{L-1}+\ket{1}\bra{0}+\ket{L-1}\bra{0}) ,
 \label{eq:H1fixedRb}
\end{align}
where we treat $H_1$ as a perturbation. 

The eigenstates of $H_0$ are $\ket{\phi_0}=\ket{0}$, forming the bound state, and $\ket{\phi_k}=\sqrt{2/L}\sum_n\sin(\pi k n /L)\ket{n}$ which are the scattering states. The corresponding unperturbed eigenenergies are given by 
$ E^{(0)}_0 =-U_{\rm BF}$ and $E^{(0)}_k =2J\ind{F}\cos(\pi k/L)$. 

The effect of $H_1$ on the eigenenergies is non-vanishing only at second order.
The second-order perturbative corrections to the energies are given by
\begin{equation}
\begin{aligned}
 \Delta E^{(2)}_k\! &=\sum_l \frac{\left|\bra{\phi_k} H_1 \ket{\phi_l}\right|^2}{E^{(0)}_k-E^{(0)}_l} \\
 &=\left\{
 \begin{array}{ll}
   \frac{8 J_{\rm F} ^2}{L}\sum_{l\text{ odd}}\frac{-\sin^2(\pi l/L)}{U_{\rm BF} +2J\ind{F}\cos(\pi l/L)} & \text{if }k=0\\
   \frac{8 J_{\rm F} ^2}{L}\frac{\sin^2(\pi k/L)}{2J_{\rm F} \cos(\pi k/L)+U_{\rm BF} } & \text{if }k\neq0, k\text{ odd} \\
   0 & \text{else.}
 \end{array}
\right.
\end{aligned}
\end{equation}
The first order corrections to the eigenstates are given by
\begin{equation}
 \begin{aligned}
  &\ket{\Delta \phi^{(1)}_k} =\sum_l \ket{\phi_l} \frac{\bra{\phi_l} H_1 \ket{\phi_k}}{E^{(0)}_k-E^{(0)}_l}\\  
  &=\left\{
  \begin{array}{ll}
    2J\ind{F}\sqrt{2/L}\sum_{l \text{ odd}}\frac{-\sin(\pi l/L)}{U\ind{BF}+2J\ind{F}\cos(\pi k/L)}\ket{l} & \text{if }k=0\\
    2J\ind{F}\sqrt{2/L}\frac{\sin(\pi k/L)}{2J\ind{F}\cos(\pi k/L)+U\ind{BF}}\ket{0} & \text{if }k \text{ odd}\\
    0 & \text{else.}
  \end{array}
\right.
 \end{aligned}
\end{equation}
With this we find that the remaining doublon fraction is given by
\begin{equation}
\begin{aligned}
  &P_2(t)=|\bra{\psi(0)}\psi(t)\rangle|^2 =\left|\sum_k e^{-i E^{(2)}_k t} \left|\mgbraket{\phi^{(1)}_k}{0}\right|^2 \right|^2 \\
  &= \left|\frac{e^{-iE_0^{(2)} t}}{N_0^2} +\sum_{k\text{ odd}}\frac{2 e^{-i E_k^{(2)}t}}{LN_k^2}\left|\frac{2J\ind{F}\sin(\pi k/L)}{2J\ind{F}\cos(\pi k/L)+U\ind{BF}} \right|^2 \right|^2
\end{aligned}
\end{equation}
where $N_0$ and $N_k$ are normalization factors. In the limit $U_{\rm BF}\gg 2J_{\rm F}$, we can approximate $2J\ind{F}\cos(\pi k/L)+U\ind{BF}\approx U\ind{BF}$ in the denominator. Neglecting terms of order $L^{-1}$ this leads to $N_0= 1+2J\ind{F}^2/U\ind{BF}^2$, $N_k= 1$, $E_0^{(2)}= -U_{\rm BF} -2J_{\rm F} ^2/U_{\rm BF}$, and $E_k^{(2)}\approx E_k^{(0)}= 2J_{\rm F}\cos(\pi k/L)$. With this we find
\begin{equation}
\begin{aligned}
  P_2(t) \approx &\left| \frac{e^{-i E_0^{(2)} t}}{1+2J\ind{F}^2/U\ind{BF}^2}\! +\! \frac{8J\ind{F}^2}{LU\ind{BF}^2}\sum_{k\text{ odd}}\!e^{-i E_k^{(0)}t} \sin^2(\pi k/L)\right|^2 \\
  \approx\, & P\ind{sat}\left( 1+ \frac{16J\ind{F}^2}{LU\ind{BF}^2} \sum_{k\text{ odd}} \sin^2(\pi k/L) \cos(\Delta E_k t)\right) \\
  \approx\, & P\ind{sat}\left[ 1 + \frac{4J_{\rm F} }{tU_{\rm BF}^2} J_1(2J_{\rm F} t)\cos\left(t\sqrt{4J_{\rm F} ^2+U_{\rm BF}^2}\right) \right]
\end{aligned}
\label{eq:P2final}
\end{equation}
where $\Delta E_k=E_0^{(2)}-E_k^{(0)}=-U\ind{BF}-2J\ind{F}^2/U\ind{BF}-2J\ind{F}\cos(\pi k/L)$ and $P\ind{sat}=1/(1+4J\ind{F}^2/U\ind{BF}^2)$.
This is the perturbative solution shown as a dashed line in Fig.~\ref{fig:one_two_doublon}(a) and (b).

\subsubsection{Perturbation Method: Relative Motion}

We now include the motion of the boson ($J\ind{B}>0$).
We first consider the non-interacting Hamiltonian describing the motion of the boson and fermion in the lattice. We denote the position of the boson in the lattice with $x_B$ and the position of the fermion with $x_F$. The non-interacting part of the Hamiltonian thus reads
\begin{equation}
    H_{\rm free} = - \sum_{\alpha={B,F}} \sum_{x_\alpha=1}^L J_\alpha \left(\vert x_\alpha \rangle \langle x_\alpha+1\vert + \vert x_\alpha \rangle \langle x_\alpha-1\vert\right)
\end{equation}

We now introduce relative, $r=x_B-x_F$, and center-of-mass coordinates, $R=\left(x_B+x_F\right)/2$, a well as total and relative quasimomenta denoted by $K$ and $k$, respectively. Then, using the ansatz $\langle R, r\vert\Psi\rangle= \sqrt{1/2\pi} e^{i K R} \psi_K(r)$ for the relative wave function at a given momentum $K$ of the center-of-mass motion (with the obvious labeling $\ket{R,r}$ of the position basis states) we find 
\begin{align*}
\langle R, r \vert & H_{\rm free}\vert \Psi\rangle \\
= &-J_{B}e^{i K R}\left(e^{i K/2 } \psi_K(r+1)+ e^{-i K/2 } \psi_K(r-1) \right)\\
&-J_{F}e^{i K R} \left(e^{-i K/2 } \psi_K(r+1)+ e^{i K/2 } \psi_K(r-1) \right)
\end{align*}

The action of $H_{\rm free}$ on the relative coordinate wavefunction $\psi_K(r)$ can be written more compactly using the quantities $J_{\pm}= \frac{J_F\pm J_B}{2}$: 
$$H_{\rm free} \psi_K(r)=\epsilon_K(k) \psi_K(r)$$
with $\epsilon_K(k)=E_K \cos\left(\phi_K+k\right)$. Here we have introduced $E_K=-4 \sqrt{J_+^2 \cos^2\left(K/2\right) + J_-^2 \sin^2\left(K/2\right)}$ and $\phi_K= \arctan\left(\frac{J_-}{J_+}\tan \left(K/2\right)\right)$.

We can now rewrite the Hamiltonian considered above for the stationary defect, interpreting the basis states $\lbrace \vert n \rangle \rbrace$ referring to the relative distance between the fermion and the boson, with $J_F\to E_K/2$. Applying a perturabtive treatment valid for $U\ind{BF}\gg J\ind{B},J\ind{F}$ results in a refined expression $P_{\rm sat}\to U_{\rm BF}^2/\left(U_{\rm BF}^2 + 8 (J_+^2+ J_-^2 )\right)$.

\subsection{Clusters of two and three doublons}
\label{sec:from_more}

\begin{figure}[t]
\centering
\includegraphics[width=\columnwidth]{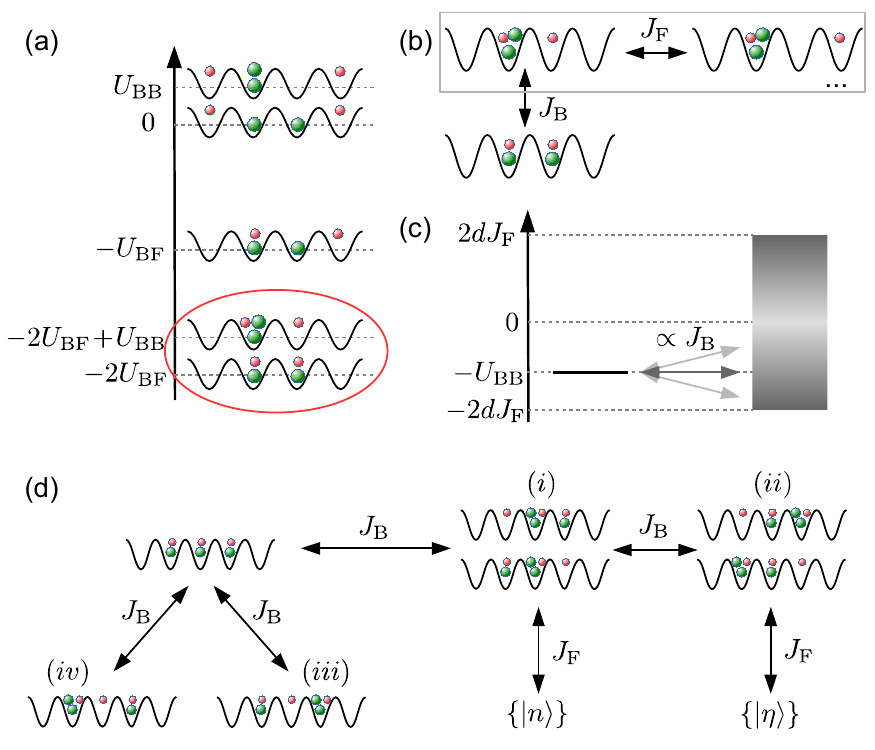}
 \caption{(a) Energy diagram for states accessible to a two-doublon cluster. For large $U\ind{BF}$ the state space can be restricted to the lowest two manifolds as indicated by the red ellipse. (b) These states are connected through different terms in the Hamiltonian. Typically the coupling $J\ind{F}$ is large compared to $J\ind{B}$. In this case we diagonalize the singlon-triplon subspace (gray box). (c) The resulting problem is a single state (two-doublon state) coupled to a continuum of singlon-triplon states. The gray shading indicates the density of states, which is minimal in the band center in 1d but maximal in the band center in higher dimensions. We have shifted the origin of the energy to be at the center of the band. (d) Relevant states and coupling matrix elements for the three-doublon problem.}
 \label{fig:23schematic}
\end{figure}

In the following we continue to assume that $U\ind{BF}$ is the largest energy scale such that we can restrict the Hilbert space to the manifold containing no lone bosons, indicated by the red ellipse in Fig.~\ref{fig:23schematic}(a).  
Consider a chain of $2L-2$ sites, where the sites are indexed by integers in the intervals $\left[-L+1, -1\right]$ and $\left[1, L-1\right]$ (omitting $0$ for symmetry). We denote the initial state in which the two doublons are placed on the nearest neighboring sites -1 and 1 by $\ket{0}$. The rest of the basis states are denoted by $\ket{n}$, $n=1,\dots,L-1$. These states are symmetrized states where the triplon and the fermion are separated by $n$ sites, cf.\ Fig.~\ref{fig:23schematic}(b). 
Using this basis and in the limit of vanishing $J_{\rm D}$ the Hamiltonian takes the form 
\begin{align}
 H&=H_0+H_1\\
 H_0&=-U_{\rm BB} \ket{0}\bra{0}+J\ind{F}\sum_{n=1}^{L-2}(\ket{n}\bra{n+1}+\ket{n+1}\bra{n})\\
 H_1&=2J_{\rm B} (\ket{0}\bra{1}+\ket{1}\bra{0}).
\end{align}

We observe that by replacing $U\ind{BB}\rightarrow U\ind{BF}$ in $H_0$ and $J\ind{B}\rightarrow J\ind{F}/2 $ in $H_1$ one obtains the single doublon Hamiltonian discussed above (except for the periodic boundaries).
Thus the problem is reduced to a single state $\ket{\phi_0}=\ket{0}$ at energy $-U\ind{BB}$ coupling to a band of states $\ket{\phi_k}=\sqrt{2/L}\sum_n\sin(\pi k n /L)\ket{n}$ of width $2J\ind{F}$ [see Fig.~\ref{fig:23schematic}(c)].

\subsubsection{Perturbative treatment}

If the the bound state and the band of scattering states are energetically separated ($U\ind{BB}>2J\ind{F}$), we can again employ the perturbative approach.
The perturbative corrections to the eigenenergies and eigenfunctions of $H_0$ are given by
\begin{equation}
 \Delta E^{(2)}_k
 = \left\{
 \begin{array}{ll}
  \frac{8 J_{\rm B} ^2}{L}\sum_{l}\frac{\sin^2(\pi l/L)}{-U_{\rm BB} -2J_{F}\cos(\pi l/L)} & \text{if }k=0\\
  \frac{8 J_{\rm B} ^2}{L}\frac{\sin^2(\pi k/L)}{2J_{K}\cos(\pi k/L)+U_{\rm BB} } & \text{if }k\neq0
 \end{array}
 \right.
\end{equation}
and 
\begin{equation}
 \begin{aligned}
  &\ket{\Delta \phi^{(1)}_k}=\\
  & \left\{ 
  \begin{array}{ll}
   2J_{\rm B} \sqrt{2/L}\sum_{l}\frac{\sin(\pi l/L)}{-U_{\rm BB} -2J\ind{F}\cos(\pi k/L)}\ket{l} & \text{if }k=0\\
   2J_{\rm B} \sqrt{2/L}\frac{\sin(\pi k/L)}{2J\ind{F}\cos(\pi k/L)+U_{\rm BB} }\ket{0} & \text{if }k \neq 0
  \end{array}
\right.
 \end{aligned}
\end{equation}

Using these expressions we find that the doublon fraction is given by 

\begin{equation}
\begin{aligned}
 P_2(t)=& \frac{ 1+ \frac{16J_{\rm B} ^2}{L} \sum_{k} \frac{\sin^2(\pi k/L)}{(U_{\rm BB} +2J\ind{F} \cos(\pi k/L))^2} \cos(\Delta E_k t)}{1+ \frac{16J_{\rm B} ^2}{L} \sum_{k} \frac{\sin^2(\pi k/L)}{(U_{\rm BB} +2J\ind{F} \cos(\pi k/L))^2}} \\
  \approx & P\ind{sat,2}\left( 1+ \frac{16J_{\rm B} ^2}{LU_{\rm BB} ^2} \sum_{k}\sin^2(\pi k/L)\cos(\Delta E_k t)\right)
\end{aligned}
\label{eq:P22final}
\end{equation}
with $\Delta E_k=-U_{\rm BB} -4J_{\rm B} ^2/U_{\rm BB} -2J\ind{F}\cos(\pi k/L)$ and $P\ind{sat,2}=1/(1+8J_{\rm B} ^2/U_{\rm BB} ^2)$, where we have replaced the denominator in the energy corrections by $U_{\rm BB}$. 

In the following we will extend the perturbative method to \emph{three doublons}. We consider a chain of $2L-3$ sites. The initial state consists of three doublons in three neighboring sites, denoted by $\ket{0}$. This state connects to two disconnected manifolds of states. The first set of states are labeled as $\ket{n}$ with $n=1, \cdots, L-1$, where $n$ labels the distance between the fermion and the triplon, and are shown in Fig.~\ref{fig:23schematic}(d) panel $(i)$. These states are connected to the states $\ket{\eta}$ (see panel $(ii)$ in  Fig.~\ref{fig:23schematic}(d)) if the triplon and the doublon on its neighboring site exchange places. For these states $\eta=1,\cdots, L-1$ labels the distance between the fermion and the doublon. We use  $\ket{n}$ and $\ket{\eta}$ to denote the symmetric combination of these two families of states. The second energetic manifold is given by the state $\ket{L_{\rm s}}$, the symmetric combination of the states $(iii)$ and $(iv)$ in Fig.~\ref{fig:23schematic}(d). In this manifold the fermionic atom cannot move and thus these states, along with $\ket{0}$ form bound states. 

In analogy with the previous perturbative treatments, we begin by determining the eigenenergies and eigenfunctions of $H_0$. The unperturbed localized states are $\vert \phi_0\rangle =\ket{0}$ and $\ket{\phi_{\rm L}}= \ket{L_{\rm s}}$ with $E_0=-U_{\rm BB}$ and $E_{\rm L}=0$. respectively. The unperturbed scattering states and their corresponding energies are given by 
\begin{align}
\vert \phi_k^\pm \rangle &= \sum_n \sqrt{\frac{2}{L}} \sin \left(\frac{\pi k}{L}n\right) \vert n \pm \rangle\\
E_k^\pm &= 2J_{\rm F} \cos\left( \frac{ \pi k}{L-1}\right)
\end{align}
where $\vert n^\pm\rangle=\frac{1}{\sqrt 2} \left(\vert n\rangle \pm \vert \eta \rangle\right)$, $n =1, \dots , L-1$. 
We now proceed to find the perturbative corrections to the wavefunctions and the energies with the perturbation Hamiltonian 

$$
H_1=2 J_B \frac{1}{\sqrt 2} \left (\vert 1^+\rangle + \vert 1^-\rangle \right)\langle 0 \vert + 2 J_B \vert \phi_L \rangle\langle 0 \vert +h.c. 
$$

To first order the eigenfunctions are given by

\begin{align}
\nonumber \vert \phi_0^{(1)} \rangle &=  -\sqrt{\frac{4}{L}} J\ind{B}\sum_{l=1}\biggl(\frac{\sin\left(\frac {\pi l}{L}\right)}{U_{\rm BB} +E_l^+}\vert \phi_l^ +\rangle \\
&+ \frac{\sin\left(\frac {\pi l}{L}\right)}{U_{\rm BB} +E_l^-}\vert \phi_l^-\rangle \biggr)- 2J_{\rm B} \frac{1}{U_{\rm BB} }\ket{\phi_{\rm L}}\\
\vert \phi_{l, \pm}^{(1)} \rangle &=  \sqrt{\frac{4}{L}} J\ind{B}\frac{\sin\left(\frac {\pi l}{L}\right)}{U_{\rm BB} +E_l^\pm}\vert \phi_0\rangle \\
\vert \phi_{\rm L }^{(1)} \rangle &= 2J_{\rm B} \frac{1}{U_{\rm BB} }\ket{\phi_0}
\end{align}
and the second order energy corrections are
\begin{align}
\nonumber E_0^{(2)}&=  \frac{4J_{\rm B}^2}{L}\left( \sum_l \frac{\sin^2\left(\frac{\pi l}{L}\right)}{-U_{\rm BB} -E_k^+}+ \sum_l\frac{\sin^2\left(\frac{\pi l}{L}\right)}{-U_{\rm BB} -E_k^-}\right)\\
&\hskip 20 pt - \frac{4 J_{\rm B}^2}{U_{\rm BB} }\approx -12 \frac{J_{\rm B}^2}{U_{\rm BB} } \\
E_{l, \pm}^{(2)}&=  \frac{4 J_{\rm B}^2}{L} \frac{\sin^2\left(\frac{\pi l}{L}\right)}{U_{\rm BB}+E_l^\pm} \approx \frac{4 J_B^2}{L} \frac{\sin^2\left(\frac{\pi l}{L}\right)}{U_{\rm BB}}\\
E_{\rm L}^{(2)}&= \frac{4 J_{\rm B}^2}{U_{\rm BB} }
\end{align}
We can now proceed as usual to find the time-dependence of the doublon fraction: 
\begin{widetext}
\begin{align}
\nonumber P_2(t)= \biggl[ 1&+\frac{8 J_{\rm B}^2}{L}\left(\sum_l \sin^2\left(\frac{\pi l}{L}\right) \left( \frac{\cos(\Delta_l^+ t)}{(U_{\rm BB}+E_l^+)^2}+\frac{\cos( \Delta_l^- t)}{(U_{\rm BB}+E_l^-)^2}\right) \right)\\
&+\frac{8 J_{\rm B}^2}{U_{\rm BB}^2} \cos(\Delta_L t)\biggr]\bigg/\left(1+\frac{8 J_{\rm B}^2}{L}\sum_l\left( \frac{\sin^2\left(\frac{\pi l}{L}\right)}{(U_{\rm BB}+E_l^+)^2}+\frac{ \sin^2\left(\frac{\pi l}{L}\right)}{(U_{\rm BB}+E_l^-)^2}\right)+\frac{8J_{\rm B}^2}{U_{\rm BB^2}}\right)
\end{align}
where 
$\Delta_l^\pm= U_{\rm BB}+\frac{12 J_{\rm B}^2}{U_{\rm BB}}-E_l^\pm$ and $\Delta_L=U_{\rm BB}+ \frac{4J_{\rm B}^2} {U_{\rm BB}^2}$. The above expression saturates to $P_{\rm sat}= 1/ \left(1+ 16 J_B^2/U_{\rm BB}^2\right)$ where we have used the approximation $U_{\rm BB}+ E_l^\pm\approx U_{\rm BB}$. 

\end{widetext}

\subsubsection{Quasi-continuum model}

We now examine the decay of two neighboring doublons in the regime $U\ind{BB}<2J\ind{F}$, i.e., when the two-doublon state lies inside the continuum of singlon-triplon states, see Fig.~\ref{fig:23schematic}(c). In this case perturbation theory breaks down since the spacings between the levels of the singlon-triplon band are small compared to the perturbation $H_1\sim J\ind{B}$.
We expect an exponential decay of the two-doublon state population and our objective here is to find its decay constant. We will use a treatment known as Bixon-Jortner quasi-continuum in quantum optics \cite{Bixon1968}. We use the notation defined in the previous section.

The Hamiltonian in the basis of bound and scattering states defined above reads
\begin{equation}
\begin{aligned}
 H&=2J\ind{B}\sqrt{2/L}\sum_{k=1}^{L-1}[\sin(\pi k/L)\ket{0}\bra{\phi_k}+h.c.] \\
   &+ \sum_{k=1}^{L-1}[2J\ind{F}\cos(\pi k/L)+U\ind{BB}]\ket{\phi_k}\bra{\phi_k} \\
  &=\sum_{k=1}^{L-1} W_k (\ket{0}\bra{\phi_k}+h.c.) + \Delta_k \ket{\phi_k}\bra{\phi_k}
  \end{aligned}
\end{equation}
Note that the couplings $W_k$ and detunings $\Delta_k$ are real-valued.
The time-dependent Schr\"odinger equation becomes
\begin{equation}
\begin{aligned}
 \dot{b}(t)&=-i\sum_{k} W_k c_k(t) \\
 \dot{c}_k(t)&=-i\Delta_k c_k(t) -iW_k b(t)
  \end{aligned}
\end{equation}
where $b(t)$ and $c_k(t)$ are the time-dependent amplitudes of states $\ket{0}$ and $\ket{\phi_k}$, respectively. Initially the system is in state $\ket{0}$, thus $b(0)=1$ and $c_k(0)=0$.
Applying a Laplace transform [$\bar{f}(s)=\mathcal{L}[f(t)]=\int_0^\infty e^{-st}f(t) dt$] this set of equations becomes
\begin{equation}
\begin{aligned}
 s\bar{b}(s)-1&=-i\sum_{k} W_k \bar{c}_k(s) \\
 s\bar{c}_k(s)&=-i\Delta_k \bar{c}_k(t) -iW_k \bar{b}(s)
  \end{aligned}
\end{equation}
Solving the second equation for $\bar{c}_k(s)$ and substituting into the first, we obtain
\begin{equation}
 \bar{b}(s)=\left(s+\sum_k \frac{W_k^2}{s+i\Delta_k} \right)^{-1}
\end{equation}
Now we have to invert the Laplace transform to obtain $b(t)$.

At this point we make the following approximation: Since $J\ind{B}\ll J\ind{F}$, the state $\ket{0}$ only significantly couples to a small window within the band of states $\ket{\phi_k}$ [see Fig.\ref{fig:23schematic}(c)]. If this window is small compared to the band width $2J\ind{F}$ we can assume that the energy differences between the states $\ket{\phi_k}$ are constant (linearize the spectrum, $\Delta_k\propto k_0-k$) and that the couplings $W_k$ are constant over the relevant range of $k$'s. Moreover, for a dense (quasi-continuous) spectrum, we can assume that there is always a resonant state $k_0$ defined by $\Delta_{k_0}=2J\ind{F}\cos(\pi k_0/L)+U\ind{BB}=0$. In quantum optics this is known as the Bixon-Jortner quasi-continuum \cite{Bixon1968}. (Originally it was reported in the context of radiationless decay in molecules.) It is a good approximation in the center of the band, i.e., for $U\ind{BB}\ll 2J\ind{F}$. Near the band edges it can be problematic.
We thus have
\begin{equation}
 \begin{aligned}
  k_0 &=\frac{L}{\pi}\arccos[U\ind{BB}/(2J\ind{F})]\\
  \Delta_k &\approx \left.\frac{d\Delta_k}{dk}\right|_{k=k_0}(k-k_0)=n\Delta \\
  W_k &\approx W_{k_0}=W
 \end{aligned}
\end{equation}
where
\begin{equation}
 \begin{aligned}
  \Delta &=2J\ind{F}\frac{\pi}{L}\sin(\pi k_0/L)= 2J\ind{F}\frac{\pi}{L}\sqrt{1-\left(\frac{U\ind{BB}}{2J\ind{F}}\right)^2} \\
  W &= 2J\ind{B}\sqrt{\frac{\pi}{L}}\sin(\pi k_0/L)= 2J\ind{B}\sqrt{\frac{\pi}{L}}\sqrt{1-\left(\frac{U\ind{BB}}{2J\ind{F}}\right)^2}
 \end{aligned}
\end{equation}
With these approximations the sum in $\bar{b}(s)$ becomes
\begin{equation}
 \sum_{n=-\infty}^{\infty}\frac{W^2}{s+i n\Delta}=\frac{\pi W^2}{\Delta}\coth\left[\frac{\pi s}{\Delta}\right]
\end{equation}
In the limit of large system size $L\rightarrow \infty$, i.e., $\Delta\rightarrow 0$, we can use $\lim_{x\rightarrow\infty}\coth(x)=1$, to obtain $\bar{b}(s)=(s+\Gamma/2)^{-1}$, with $\Gamma=2\pi W^2/\Delta$, which is obviously the Laplace transform of $b(t)=\exp(-\Gamma t/2)$. Note that $W^2/\Delta$ is independent of $L$.

To summarize the final result, we have
\begin{equation}
  P_2(t)=|b(t)|^2= e^{-\Gamma t}
\end{equation}
with
\begin{equation}
 \Gamma=\frac{8J\ind{B}^2}{J\ind{F}}\sqrt{1-\left(\frac{U\ind{BB}}{2J\ind{F}}\right)^2}
\end{equation}
This model was used to produce the black dashed lines in Figs.~\ref{fig:one_two_doublon}c and d.

\subsubsection{Higher dimensions}

Consider the problem of 2 doublons on neighboring sites in $d$ dimensions ($d>1$). We again neglect all states that are detuned by at least $U\ind{BF}$ from the initial state. Analogous to the 1d case we can split the problem into the two-doublon state plus a continuum of singlon-triplon states. As in the case of a single doublon in higher dimensions, the problem is that the eigenstates of the singlon-triplon block cannot be found analytically. They are the eigenstates of a free particle on a $d$-dimensional lattice with a defect, i.e., a single site that is not accessible (the site that is occupied by the triplon). Unlike in the single doublon problem, the coupling of the bound state to the scattering states is $J\ind{B}$, which is different from the tunneling rate $J\ind{F}$ that determines the width $4dJ\ind{F}$ of the band of scattering states. Note that the triplon is in principle also mobile for $d>1$, since the fermion (singlon) can propagate to any of the $2d$ neighboring sites of the triplon and a boson can subsequently tunnel to that site to form two doublons on sites different from the initial doublon sites, which can then again decay into a triplon on the neighboring site of the original triplon site plus a singlon. This process leads to a center-of-mass motion of the cluster.

However, for the understanding of the short-time dynamics, it is sufficient to consider a single two-doublon state plus a continuum of scattering states. The global properties of the band are not changed significantly with respect to a perfect lattice by the addition of a defect, in particular the density of states and the width of the band stays the same. The crucial parameter is still the position of the two-doublon state with respect to the band: If it is outside the band ($U\ind{BB}>2dJ_F$), $P_2(t)$ will quickly saturate and $1-P\ind{sat}\propto J\ind{B}^2$, the square of the coupling matrix element. $P\ind{sat}$ will depend on $U\ind{BB}/J_F$ in a complicated way, and for $U\ind{BB}/J_F\gg 2d$ we have $1-P\ind{sat}\propto J\ind{B}^2/U\ind{B}^2$, however, in this limit $1-P\ind{sat}$ will be extremely small and the doublon pairs are essentially stable. Thus, for $J\ind{B}\ll J\ind{F}$, the doublon pairs become stable very quickly as soon as $U\ind{BB}>2dJ_F$.

In the regime $U\ind{BB}<2d J\ind{F}$, where the bound state lies inside the band of scattering states, we can again employ the Bixon-Jortner quasi-continuum approximation, which means that we assume that the bound state couples to a continuum of equally spaced states (spacing $\Delta$) with a constant coupling strength $W$. If this is a valid description of the continuum, then in the limit of large $L$, the bound state population will decay like $\exp[-\Gamma t]$, with $\Gamma=2\pi W^2/\Delta$.
In a higher-dimensional situation, for a given energy of the bound state, there are in general several resonant states ($\Delta_k=0$) and each of those states couples to the bound state with a different $W_k$. What we can do is to define $W^2$ as the average squared coupling of the scattering states at a given energy and $\Delta$ as the mean splitting between neighboring states in an energy interval, i.e., as the inverse of the density of states $\rho(E=-U\ind{BB})=\int d\mathbf{k}\delta(E(\mathbf{k})+U\ind{BB})$.
Calculating the density of states involves an elliptical integral and cannot be solved analytically. However, it is clear that the density of states will be proportional to $1/J\ind{F}$ and thus $\Delta\propto J\ind{F}$. The squared coupling will be proportional to $J\ind{B}^2$. Thus $\Gamma=(J\ind{B}^2/J\ind{F})f(u)$, where $u=U\ind{BB}/J\ind{F}$. The function $f(u)$ depends on where in the band the bound state is located ($0<u<d$) and will decrease to zero at the band edge. 

\begin{figure}[t]
  \centering
 \includegraphics[width=\columnwidth]{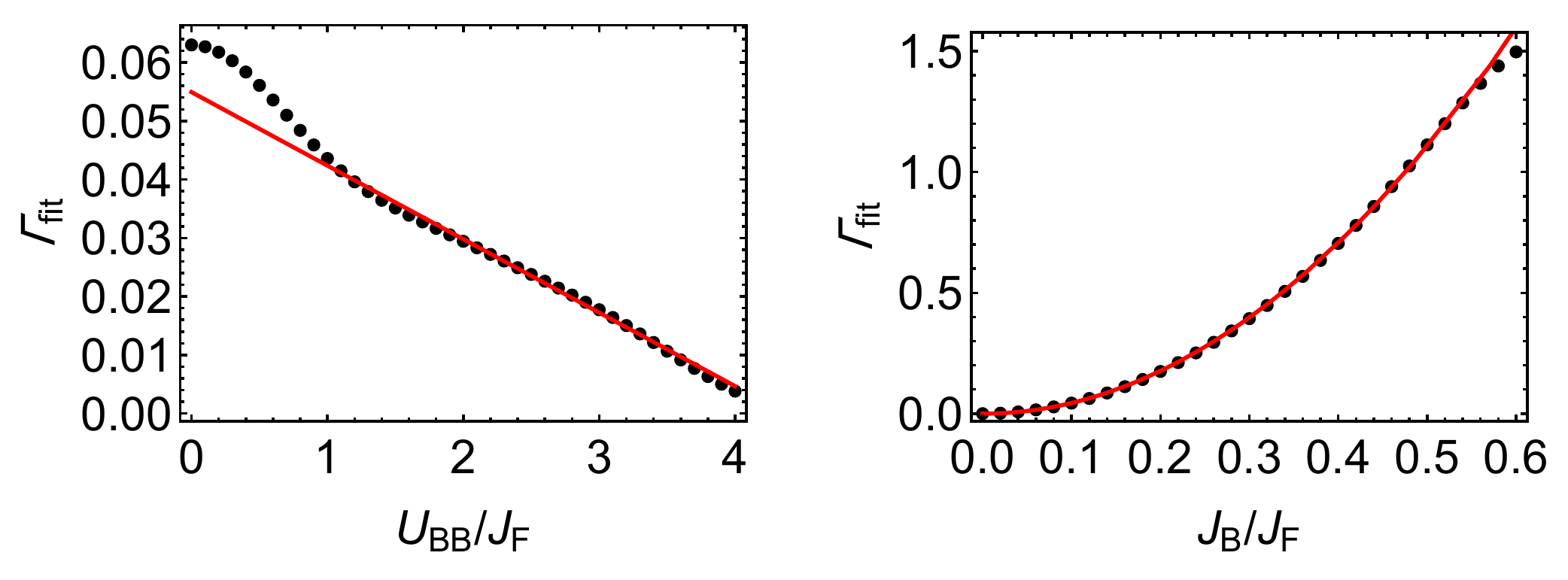}
 \caption{Doublon decay in 2d. Fitted decay rate $\Gamma$ from fitting $P_2(t)$ with $\exp[-\Gamma t]$ as a function of $U\ind{BB}/J\ind{F}$ at $J\ind{B}/J\ind{F}=0.1$ (left) and of $J\ind{B}/J\ind{F}$ at $U\ind{BB}/J\ind{F}=1$ (right). Solid lines are a linear fit to the region $1<U\ind{BB}/J\ind{F}<2d$, and a quadratic fit, respectively. A lattice of size $22$ by $23$ sites was used with the doublon pair initially sitting on the two central sites.}
 \label{fig:fitGamma}
\end{figure}

In Fig.~\ref{fig:fitGamma} we show the fitted decay rates $\Gamma$, obtained by fitting the function $\exp[-\Gamma t]$ to the numerically obtained $P_2(t)$ as a function of the rescaled model parameters. We find that an exponential decay is indeed a good fit except close to the edges of the band $|U\ind{BB}|\approx 2dJ\ind{F}$. We observe that $\Gamma$ decreases to zero monotonically in the interval $0<U\ind{BB}/J\ind{F}<2d$ and is approximately linear for $U\ind{BB}/J\ind{F}>1$. In addition, $\Gamma\propto J\ind{B}^2/J\ind{F}$ is very well satisfied for $J\ind{B}/J\ind{F}\lesssim 0.5$. These statements also apply for $d=3$.

\begin{acknowledgments}
We thank
Jun Ye, Jacob Covey, Steven Moses, and Michael Wall
for helpful discussions. 
This work is supported by the AFOSR grant FA9550-18-1-0319 and its MURI Initiative, the ARO single investigator award W911NF-19-1-0210,  the NSF PHY1820885, NSF JILA-PFC PHY-1734006 grants, and by NIST.
M.G.~acknowledges support from the DFG Collaborative Research Center SFB1225 (ISOQUANT). J.S.~is supported by the French National Research Agency (ANR) through the Programme d'Investissement d'Avenir under contract ANR-11-LABX-0058\_NIE within the Investissement d'Avenir program ANR-10-IDEX-0002-02.
\end{acknowledgments}

% \bibliography{doublon_refs}{} 

%merlin.mbs apsrev4-1.bst 2010-07-25 4.21a (PWD, AO, DPC) hacked
%Control: key (0)
%Control: author (8) initials jnrlst
%Control: editor formatted (1) identically to author
%Control: production of article title (-1) disabled
%Control: page (0) single
%Control: year (1) truncated
%Control: production of eprint (0) enabled
%

\end{document}